\documentclass[conference]{IEEEtran}

\usepackage{amsmath, amssymb, bm, cite, epsfig, psfrag}
\usepackage{graphicx}
\usepackage{array}
\usepackage{lipsum}
\usepackage{ragged2e}
\newcolumntype{P}[1]{>{\centering\hspace{0pt}}p{#1}}
\newcolumntype{M}[1]{>{\centering\hspace{0pt}}m{#1}}
\newcolumntype{L}{>{\centering\arraybackslash}m{3cm}}
\usepackage[margin=10pt,font=small]{caption}
\usepackage{epstopdf}

\usepackage{bbm}
\usepackage{multirow}
\usepackage[usenames,dvipsnames]{xcolor}
\renewcommand{\arraystretch}{1.5}
\usepackage{subfigure}
\usepackage{etoolbox}
\usepackage{pbox}
\usepackage[width=0.48\textwidth]{caption}
\usepackage{colortbl}
\usepackage{fancyhdr}
\usepackage{bm}
\newtoggle{conference}
\togglefalse{conference} 
\graphicspath{{figures/}}

\def\taubar{\overline{\tau}}

\def\taubar{\overline{\tau}}

\def\PL{\mathrm{PL}}

\def\tm1{t\! - \! 1}
\def\tp1{t\! + \! 1}

\fancypagestyle{plain}{
  \fancyhf{}
  \fancyhead[C]{Conference on \LaTeX}     
  \fancyfoot[L]{This is a notice}

}
\usepackage{tikz}
\usetikzlibrary{calc}
\begin{document}

\bibliographystyle{IEEEtran}

\title{Exploiting Directionality for Millimeter-Wave Wireless System Improvement}

\author{\IEEEauthorblockN{George R. MacCartney Jr., Mathew K. Samimi, and Theodore S. Rappaport}
\IEEEauthorblockA{NYU WIRELESS\\
NYU Polytechnic School of Engineering\\
Brooklyn, NY 11201\\
\{gmac,mks,tsr\}@nyu.edu}
}


\maketitle
\begin{tikzpicture}[remember picture, overlay]
\node at ($(current page.north) + (0,-0.25in)$) {G. R. MacCartney, Jr., M. K. Samimi, and T. S. Rappaport, ``Exploiting Directionality for Millimeter-Wave Wireless System};
\node at ($(current page.north) + (0,-0.4in)$) {Improvement," accepted in \textit{2015 IEEE International Conference on Communications (ICC)}, June 2015.};
\end{tikzpicture}
\thispagestyle{empty}
\begin{abstract}
This paper presents directional and omnidirectional RMS delay spread statistics obtained from 28 GHz and 73 GHz ultrawideband propagation measurements carried out in New York City using a 400 Megachips per second broadband sliding correlator channel sounder and highly directional steerable horn antennas. The 28 GHz measurements did not systematically seek the optimum antenna pointing angles and resulted in 33\% outage for 39 T-R separation distances within 200 m. The 73 GHz measurements systematically found the best antenna pointing angles and resulted in 14.3\% outage for 35 T-R separation distances within 200 m, all for mobile height receivers. Pointing the antennas to yield the strongest received power is shown to significantly reduce RMS delay spreads in line-of-sight (LOS) environments. A new term, \textit{distance extension exponent} (DEE) is defined, and used to mathematically describe the increase in coverage distance that results by combining beams from angles with the strongest received power at a given location. These results suggest that employing directionality in millimeter-wave communications systems will reduce inter-symbol interference, improve link margin at cell edges, and enhance overall system performance.
\end{abstract}


\section{Introduction}~\label{sec:intro}
The growing demand for wireless broadband communications has led to the exploration of the underutilized millimeter-wave (mmWave) spectrum where a vast amount of raw bandwidth can be exploited to provide multi-gigabit per second data rates for backhaul, fronthaul, and mobile applications~\cite{Pi:CommMag11}. Highly directional horn antennas at the transmitter (TX) and receiver (RX) make up for the additional free space path loss induced by the order of increase in carrier frequency, resulting in many more directional systems than at today's Microwave and Ultra-High Frequency (UHF) bands below 6 GHz, where quasi-omnidirectional antennas are commonplace~\cite{TSR:mmWave}. 
\footnotetext[1]{Three additional measurement studies in 2012 at 28 GHz were conducted in Manhattan and Brooklyn but are not considered for this study since the scenarios were different, including:  (1) 28 GHz measurements with a narrowbeam TX antenna (10.9$^\circ$ HPBW) and widebeam RX antennas (28.8$^\circ$ HPBW) were conducted for 5 RX locations in Manhattan. (2) 28 GHz measurements with narrowbeam TX and RX antennas (10.9$^\circ$ HPBW) were conducted for 5 RX locations in Brooklyn. (3) 28 GHz measurements with a widebeam TX antenna (28.8$^\circ$ HPBW) and narrowbeam RX antennas (10.9$^\circ$ HPBW) were conducted for 4 RX locations in Brooklyn~\cite{SR:IMS14}.}
Recent outdoor mmWave measurements using less than 1 W of transmit power have shown that future cell radii will be 200 m or so, implying that at such short distances, atmospheric and rain attenuations will not be a major concern for mmWave outdoor urban-microcell (UMi) propagation~\cite{TSR:WillWork13}. 
\begin{table}[t!]
\renewcommand{\arraystretch}{1.15}
\begin{center}
\caption{T-R separation distances and number of TX-RX location combinations where signal was detected, and where outages occurred, for LOS and NLOS environments at 28 GHz (2012) and 73 GHz in Manhattan (2013). Note that the 28 GHz campaign did not systematically search for pointing angles between the TX and RX antennas to make a link, but the 73 GHz campaign did.}~\label{tbl:outage}
\begin{tabular}{|>{\centering\arraybackslash}m{1.6cm}|>{\centering\arraybackslash}m{0.62cm}|>{\centering\arraybackslash}m{2.7cm}|>{\centering\arraybackslash}m{2.62cm}|}
\hline
\multicolumn{4}{|c|}{\textbf{Manhattan Measurements\footnotemark[1]}}  \tabularnewline \hline
\multicolumn{2}{|c|}{} 													& \textbf{28 GHz} 								& \textbf{73 GHz}  								\tabularnewline 
\multicolumn{2}{|c|}{} 													& \textbf{TX/RX: 10.9$^\circ$ HPBW}				& \textbf{TX/RX: 7$^\circ$ HPBW} 				\tabularnewline \hline
{\textbf{\# of}} & \multirow{2}{0.62cm}{\centering{\textbf{LOS}}}		& \multirow{2}{2.7cm}{\centering{\textbf{6}}}	& \multirow{2}{2.7cm}{\centering{\textbf{5}}}	\tabularnewline 
{\textbf{locations}} 			& 													& 												&												\tabularnewline \cline{2-4}
{\textbf{measured}}  			& \multirow{2}{0.62cm}{\centering{\textbf{NLOS}}} 	& \multirow{2}{2.7cm}{\centering{\textbf{33}}}	& \multirow{2}{2.7cm}{\centering{\textbf{30}}}	\tabularnewline 
{\textbf{($\bm{d} \leq$ 200 m)}}& 													& 												&  \tabularnewline \hline
{\textbf{\# of}} 				& \multirow{2}{0.62cm}{\centering{\textbf{LOS}}}	& \multirow{2}{2.7cm}{\centering{\textbf{6}}}	& \multirow{2}{2.7cm}{\centering{\textbf{5}}}	\tabularnewline 
{\textbf{locations}} 			& 													& 												&												\tabularnewline \cline{2-4}
{\textbf{measured}}  			& \multirow{2}{0.62cm}{\centering{\textbf{NLOS}}} 	& \multirow{2}{2.7cm}{\centering{\textbf{68}}}	& \multirow{2}{2.7cm}{\centering{\textbf{31}}}	\tabularnewline 
{\textbf{for all $\bm{d}$}} 	& 													& 												& 												\tabularnewline \hline
{\textbf{\# of}} 				& \multirow{2}{0.62cm}{\centering{\textbf{LOS}}}	& \centering{\textbf{6}}						& \centering{\textbf{5}}						\tabularnewline 
{\textbf{locations}} 			& 													& (31 m $\leq d \leq$ 102 m)					& (30 m $\leq d \leq$ 54 m)						\tabularnewline \cline{2-4}
{\textbf{with signal}}			& \multirow{2}{0.62cm}{\centering{\textbf{NLOS}}}	& \centering{\textbf{20}}						& \centering{\textbf{25}}						\tabularnewline 
{\textbf{($\bm{d} \leq$ 200 m)}}& 													& (61 m $\leq d \leq$ 187 m)					& (48 m $\leq d \leq$ 190 m)					\tabularnewline \hline
{\textbf{\# of outage}} 		& {\centering{\textbf{LOS}}}						& \centering{\textbf{0}}						& \centering{\textbf{0}}						\tabularnewline \cline{2-4}
{\textbf{locations}} 			& \multirow{2}{0.62cm}{\centering{\textbf{NLOS}}}	& \centering{\textbf{13}}						& \centering{\textbf{5}}						\tabularnewline 
{\textbf{($\bm{d} \leq$ 200 m)}}& 													& (96 m $\leq d \leq$ 193 m)					& (168 m $\leq d \leq$ 198 m)					\tabularnewline \hline
{\textbf{\# of}} 				& \multirow{2}{0.62cm}{\centering{\textbf{LOS}}}	& \centering{\textbf{6}}						& \centering{\textbf{5}}						\tabularnewline 
{\textbf{locations}}			& 													& (31 m $\leq d \leq$ 102 m)					& (30 m $\leq d \leq$ 54 m)						\tabularnewline \cline{2-4}
{\textbf{with signal}}			& \multirow{2}{0.62cm}{\centering{\textbf{NLOS}}}	& \centering{\textbf{20}}						& \centering{\textbf{25}}						\tabularnewline 
{\textbf{for all $\bm{d}$}}		& 													& (61 m $\leq d \leq$ 187 m)					& (48 m $\leq d \leq$ 190 m)					\tabularnewline \hline
{\textbf{\# of outage}} 		& {\centering{\textbf{LOS}}}						& \centering{\textbf{0}}						& \centering{\textbf{0}}						\tabularnewline \cline{2-4}
{\textbf{locations}} 			& \multirow{2}{0.62cm}{\centering{\textbf{NLOS}}}	& \centering{\textbf{48}}						& \centering{\textbf{6}}						\tabularnewline 
{\textbf{for all $\bm{d}$}}		& 													& (96 m $\leq d \leq$ 425 m)					& (168 m $\leq d \leq$ 216 m)					\tabularnewline \hline
\end{tabular}
\end{center}
\end{table}

Future mmWave radio-systems must be designed appropriately with the help of statistical channel models (SCM), such as the widespread 3GPP and WINNER II models that were used to characterize the sub-6 GHz wireless channel~\cite{ICC2015:SR}. Recent mmWave propagation measurements, have faithfully accounted for mmWave wideband channel properties, with new statistical directional and omnidirectional path loss models, and temporal and spatial channel models~\cite{MR:ICC14,RBMQ:ICC12,MKR:PIMRC14,MR:GCOM14,Hur14:2,MIWEBA}.

The time dispersive characteristics of mmWave wideband channels have, however, received little attention to date. The omnidirectional root-mean-square (RMS) delay spread provides an important measure of channel time dispersion and multipath, and greatly impacts equalization, cyclic-prefixes, and receiver architectures when designing UWB systems. Multipath components (MPCs) arise from propagating signals that experience reflection and scattering, and cause inter-symbol interference (ISI). The measured RMS delay spread thus provides valuable channel knowledge. 
\begin{table}[t!]
\renewcommand{\arraystretch}{1.1}
\begin{center}
\caption{Broadband sliding correlator channel sounder system specifications used to conduct the 28 GHz and 73 GHz dense UMi propagation measurements in New York City~\cite{MR:ICC14}.}~\label{tbl:SounderSpecs}
\begin{tabular}{| >{\centering\arraybackslash}m{3cm} || >{\centering\arraybackslash}m{2cm} | >{\centering\arraybackslash}m{2cm} |}\hline
\multicolumn{3}{|c|}{\textbf{Manhattan Measurements}}  \tabularnewline \hline
\textbf{Campaign} & \textbf{28 GHz (2012)} & \textbf{73 GHz (2013)} \\ \hline
\textbf{TX / RX IF Frequency} & 5.4 GHz & 5.625 GHz \\ \hline
\textbf{TX / RX LO Frequency} & 22.6 GHz & 67.875 GHz \\ \hline
\textbf{TX / RX PN Chip Rate} & \multicolumn{2}{c|}{400 Mcps / 399.95 Mcps} \\ \hline
\textbf{RF Bandwidth (Null-to-null)} & \multicolumn{2}{c|}{800 MHz} \\ \hline
\textbf{Multipath Time Resolution} & \multicolumn{2}{c|}{2.5 ns} \\ \hline
\textbf{Max. TX RF Power} & 30 dBm & 14.6 dBm \\ \hline
\textbf{Max. Measurable Path Loss (5 dB SNR)} & 178 dB & 181 dB \\ \hline
\textbf{TX / RX Antenna Polarization} & \multicolumn{2}{c|}{Vertical} \\ \hline
\textbf{TX / RX Antenna Gain} & 24.5 dBi & 27 dBi \\ \hline
\textbf{TX / RX Az. HPBW} & 10.9$^\circ$ & 7$^\circ$ \\ \hline
\textbf{TX / RX El. HPBW} & 8.6$^\circ$ & 7$^\circ$ \\ \hline
\textbf{TX Antenna Heights} & \multicolumn{2}{c|}{7 and 17 m} \\ \hline
\textbf{RX Antenna Heights} & 1.5 m & 2 m \\ \hline
\end{tabular}
\end{center}
\end{table}
Omnidirectional RMS delay spreads were extensively measured for the UHF/Microwave bands in order to design current 3G and 4G systems, resulting in RMS delay spreads from 15.7 ns to 23.75 ns with an UWB pulse of 2.2 GHz between 3.1 and 5.3 GHz~\cite{ZNC:ICIAS14}. Typical mean RMS delay spreads at 400 MHz and 1900 MHz in multipath-rich LOS and NLOS environments were measured on the order of 300-400 ns and 730 ns, respectively, over 10-20 MHz of RF bandwidth~\cite{MLN:ISWCS08,MKPC:PIMRC02}. The mean and maximum RMS delay spreads observed using directional antennas in a NLOS dense urban wideband cellular study were 12.2 ns and 117 ns at 38 GHz, respectively~\cite{RBMQ:ICC12}. 90\% of 59 GHz collected wideband measurements had RMS delay spreads less than 20 ns in a dense UMi environment for T-R separation distances less than 200 m~\cite{Lovnes94:1}, and the typical RMS delay spread in another study in a city street environment was below 20 ns~\cite{Smulders}.

28 GHz and 73 GHz mmWave UWB propagation measurements in New York City were collected with a 400 Megachips per second (Mcps) broadband sliding correlator channel sounder, each with a pair of highly directional steerable horn antennas used at the TX and RX, for 74 and 36 TX-RX location combinations at 28 GHz and 73 GHz, respectively, for mobile RX heights~\cite{TSR:WillWork13,MR:ICC14}. Power delay profiles (PDPs) were measured at unique azimuth and elevation antenna pointing angles at the TX and RX, allowing us to model both temporal and spatial channel characteristics. Table~\ref{tbl:outage} lists the number of LOS and NLOS locations measured and the range of distances for both 28 GHz and 73 GHz LOS and NLOS environments in Manhattan, where signal was detected and where outages occurred. High-gain directional antennas were used to complete mmWave links in order to make up for the increase in free space path loss at higher frequencies compared to today's bands. MmWave channels must therefore be characterized using directional statistical  models obtained from unique pointing angle narrowbeam measurements, as reported in~\cite{TSR:WillWork13,MR:ICC14,Shu:CommMagDec14}. In this paper, we present both directional and omnidirectional RMS delay spread statistics based on our 28 GHz and 73 GHz propagation measurements, that can be used in developing beam combining and beamforming algorithms to be used in future electrically-steered on-chip antennas~\cite{RMG:PROC11}.

Section~\ref{sec:Drms} presents directional RMS delay spreads obtained from all pointing angle measurements between the TX and RX, Section~\ref{sec:CovExt} introduces a new term, the \textit{distance extension exponent} (DEE), used to determine distance extension when combining beams from unique pointing angles with the strongest received power, Section~\ref{sec:3Drms} gives omnidirectional RMS delay spreads synthesized from absolute timing power delay profiles recovered using 3-D ray-tracing techniques, and Section~\ref{sec:conc} concludes the paper. 

\section{Directional RMS Delay Spread}~\label{sec:Drms}
UWB propagation measurement campaigns were performed in 2012 and 2013 in dense urban LOS and NLOS environments to assess the viability of mmWave outdoor communications. 400 Mcps broadband sliding correlator channel sounders at 28 and 73 GHz shared a similar architecture but with different IF, LO, and  RF up- and down-conversion front-ends. T-R separation distances ranged from 30 m to 425 m, where several thousands of PDPs were recorded at many different azimuth and elevation pointing angle combinations using steerable high-gain horn antennas (10.9$^{\circ}$ (at 28 GHz) and 7$^{\circ}$ (at 73 GHz), half-power beamwidth (HPBW)) at both the TX and RX, providing maximum measurable path loss of 178 dB and 181 dB at 28 GHz and 73 GHz, respectively. Although both backhaul (4.06 m) and mobile (2 m) RX heights were measured at 73 GHz, only mobile measurements are considered in this paper for proper comparison with the 28 GHz mobile (1.5 m) measurements.

The 28 GHz measurements were more rigid, in that they did not systematically point and search for antenna pointing angles between the TX and RX that resulted in the strongest power, whereas the 73 GHz measurement campaign did search for TX and RX antenna pointing angles that resulted in links with the strongest received power. At both 28 and 73 GHz, omnidirectional path loss models were created by summing PDPs from adjacent and orthogonal antenna beams so as to not double count overlapping angles or multipath energy. While the 73 GHz measurements scanned a larger portion of the 4$\pi$ steradian sphere, both measurements resulted in consistent path loss models~\cite{MKR:PIMRC14}. The measurement equipment specifications are listed in Table~\ref{tbl:SounderSpecs}, and additional details are available in~\cite{TSR:WillWork13,MR:ICC14}.
\begin{table}[t!]
\renewcommand{\arraystretch}{1.4}
\begin{center}
\caption{LOS and NLOS mean RMS delay spreads and standard deviations in New York City at 28 GHz and 73 GHz over all arbitrary directional antenna pointing angles in the azimuth and elevation planes for all locations with detectable signal.}~\label{tbl:RMSdelay}
\begin{tabular}{|>{\centering\arraybackslash}m{1.9cm}|>{\centering\arraybackslash}m{1.9cm}|>{\centering\arraybackslash}m{0.6cm}|>{\centering\arraybackslash}m{0.6cm}|>{\centering\arraybackslash}m{0.6cm}|>{\centering\arraybackslash}m{0.6cm}|}
\hline
\multicolumn{2}{|c||}{\multirow{2}{3.8cm}{\scriptsize{\textbf{Directional RMS Delay Spread for Locations with Signal Over All Pointing Angles and Distances Between the TX and RX}}}} & \multicolumn{2}{c|}{\textbf{LOS}} & \multicolumn{2}{c|}{\textbf{NLOS}} \tabularnewline \cline{3-6}
\multicolumn{2}{|c||}{} 																					& \textbf{$\bm{\mu}$ (ns)} 	& \textbf{$\bm{\sigma}$ (ns)}	& \textbf{$\bm{\mu}$ (ns)}		& \textbf{$\bm{\sigma}$ (ns)}	\tabularnewline \hline \hline
\multicolumn{2}{|c||}{\textbf{28 GHz (10.9$^\circ$ HPBW)}}	& 28.8 							& 44.3 						& 17.4 							& 28.3 							\tabularnewline \cline{2-6} 
\multicolumn{2}{|c||}{\textbf{73 GHz (7$^\circ$ HPBW)}}		& 13.9 							& 22.0 						& 11.1 							& 22.9 							\tabularnewline \hline
\end{tabular}
\end{center}
\end{table}
\begin{figure}[h!]
	\begin{center}
		\includegraphics[width=0.5\textwidth]{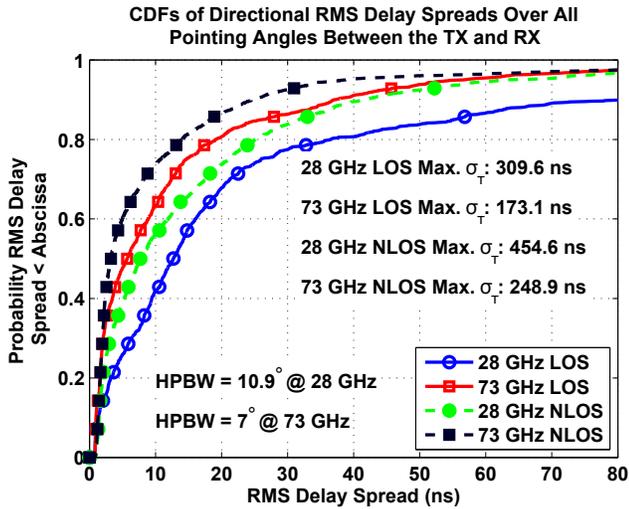}
		\caption{28 GHz and 73 GHz LOS and NLOS RMS delay spread CDFs over all pointing angles in the azimuth and elevation planes in New York City using high-gain directional antennas (HPBW of 10.9$^\circ$ at 28 GHz and 7$^\circ$ at 73 GHz). 28 GHz RMS delay spreads are larger than 73 GHz RMS delay spreads in both LOS and NLOS environments.}\label{fig:Drms}
	\end{center}
\end{figure}

Fig.~\ref{fig:Drms} shows cumulative distribution functions (CDFs) of the LOS and NLOS directional RMS delay spreads at 28 GHz and 73 GHz over all pointing angles between the TX and RX antennas, where RMS delay spreads $\sigma_{\tau}(\theta_i,\phi_j)$ at a given azimuth angle $\theta_i$ and elevation angle $\phi_j$ combination were computed from Eqs.~(\ref{eq:eq1}) to~(\ref{eq:eq3})~\cite{TSR:mmWave}:
\begin{equation}
\taubar(\theta_i,\phi_j) = \frac{\sum_k  P(\tau_k,\theta_i,\phi_j)\tau_k}{\sum_k P(\tau_k,\theta_i,\phi_j)}\label{eq:eq1}
\end{equation}
\begin{equation}
\overline{\tau^{2}}(\theta_i,\phi_j) = \frac{\sum_k  P(\tau_k,\theta_i,\phi_j){\tau_k}^2}{\sum_k P(\tau_k,\theta_i,\phi_j)}\label{eq:eq2}
\end{equation}
\begin{equation}
\sigma_\tau(\theta_i,\phi_j) = \sqrt{\overline{\tau^{2}}-(\taubar)^{2}}\label{eq:eq3}
\end{equation}
where $P(\tau_k,\theta_i,\phi_j)$ is the relative amplitude of multipath in mW/ns in time bin $\tau_k$, for a measured azimuth angle $\theta_i$ and elevation angle $\phi_j$ direction. In our work, each directional PDP measurement was processed using a 5 dB above mean thermal noise floor SNR threshold, where all measured signals above this threshold were deemed valid and all signal levels below were ignored. Directional LOS RMS delay spreads over all arbitrary pointing angles are for RX locations where the TX and RX were in clear LOS of each other, with the TX and RX antennas aligned on boresight, but also when TX and RX antennas were not necessarily aligned on boresight, in order to measure delay spread of a LOS environment with directional antennas~\cite{TSR:WillWork13,MR:ICC14}. The large RMS delay spreads observed in LOS usually appear at off-boresight angles arising from multipath signal reflections and scattering in the environment, causing large delays with strong signals. Directional NLOS RMS delay spreads are for RX locations where obstructions blocked the clear optical LOS path from the TX. Table~\ref{tbl:outage} shows the number of TX-RX location combinations where signal was measured, and the corresponding T-R separation distances for LOS and NLOS environments at 28 GHz and 73 GHz.

\begin{figure}
	\begin{center}
		\includegraphics[width=0.5\textwidth]{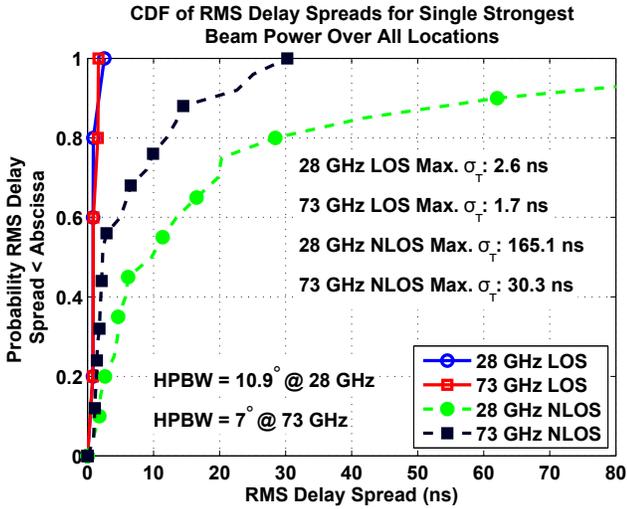}
		\caption{28 GHz and 73 GHz directional RMS delay spread CDFs for the single strongest beam power\textsuperscript{2} over all TX-RX location combinations using directional antennas (HPBW of 10.9$^\circ$ at 28 GHz and 7$^\circ$ at 73 GHz) in New York City in LOS and NLOS environments. Table~\ref{tbl:outage} shows the corresponding T-R separation distances and measured locations. Table~\ref{tbl:bestRMSdelay} show the mean RMS delay spreads and standard deviations for the strongest single beam power over all locations\textsuperscript{3}.\label{fig:Bestrms}}
	\end{center}
\end{figure}

Table~\ref{tbl:RMSdelay} provides the directional mean RMS delay spreads and standard deviations observed at 28 GHz and 73 GHz in LOS and NLOS environments. The mean LOS RMS delay spreads measured at 28 GHz and 73 GHz were 28.8 ns and 13.9 ns, respectively, and are both larger than their corresponding mean NLOS RMS delay spreads of 17.4 ns and 11.1 ns, indicating that scattered energy is more prominent in LOS environments when looking in arbitrary directions. Fig.~\ref{fig:Drms} displays that 90\% of the measured RMS delay spreads occur within 80 ns and 38 ns in LOS environments at 28 GHz and 73 GHz, respectively, whereas 90\% of measured RMS delay spreads in NLOS environments are within 39 ns and 25 ns at 28 GHz and 73 GHz, respectively. The larger LOS and NLOS RMS delay spreads at 28 GHz imply that reflections are stronger at 28 GHz and that diffuse scattering is more important for the propagation path at 73 GHz (this is also seen in Figs.~\ref{fig:Drms} and~\ref{fig:Bestrms}, and Tables~\ref{tbl:RMSdelay} and~\ref{tbl:bestRMSdelay}). The NLOS mean RMS delay spreads at 28 GHz and 73 GHz compare well with other mmWave RMS delay spreads that are typically lower than 20 ns~\cite{Smulders,Lovnes94:1}.

Large RMS delay spreads in LOS environments occur when the TX and RX antennas are not aligned on boresight. Proper beam pointing in mmWave systems can direct antennas to the strongest received power or smallest delay spread angle combinations between TX and RX antennas, effectively reducing RMS delay spread and resulting in an improved link budget~\cite{Shu:CommMagDec14}.
Fig.~\ref{fig:Bestrms} shows the RMS delay spread CDFs from the unique pointing angles (in both azimuth and elevation lane) at both the TX and RX that resulted in the strongest received power for each TX-RX location combination in LOS and NLOS environments.  Fig.~\ref{fig:Bestrms} reveals that directional RMS delay spreads when only considering the single strongest beam power for a TX-RX location combination, are less than 2 ns at 28 and 73 GHz, thus illustrating that the angles with the strongest received power correspond to the LOS boresight link. Table~\ref{tbl:bestRMSdelay} shows the directional single strongest beam LOS and NLOS mean RMS delay spreads and standard deviations at 28 GHz and 73 GHz. Improvements for the mean RMS delay spread when considering the single strongest beam powers compared to arbitrary beams, are observed in LOS environments, where the strongest path is always the boresight-to-boresight link between the TX and RX antennas, compared to any arbitrary direction. At 28 GHz, the single strongest beam power mean RMS delay spread over all locations for NLOS is larger than for the arbitrary pointing angle case due to a large number of angles at 28 GHz with low RMS delay spreads, but when comparing Fig.~\ref{fig:Drms} and Fig.~\ref{fig:Bestrms}, the trend of the CDF curve is relatively consistent up until the 70\% mark. The higher 28 GHz delay spread is likely due to the experimental design of the 28 GHz measurements that did not seek out the stongest beam pointing angles at the TX and RX. With beam searching and beamforming algorithms, directional RMS delay spreads for the strongest single beam powers will reduce ISI, thereby significantly improving connectivity and data throughput in a LOS mmWave network~\cite{Shu:CommMagDec14}.
\begin{table}[t!]
\renewcommand{\arraystretch}{1.2}
\begin{center}
\caption{LOS and NLOS mean RMS delay spreads and standard deviations for the strongest single beam power over all locations in New York City at 28 GHz and 73 GHz.}~\label{tbl:bestRMSdelay}
\begin{tabular}{|>{\centering\arraybackslash}m{1.9cm}|>{\centering\arraybackslash}m{1.9cm}|>{\centering\arraybackslash}m{0.6cm}|>{\centering\arraybackslash}m{0.6cm}|>{\centering\arraybackslash}m{0.6cm}|>{\centering\arraybackslash}m{0.6cm}|}
\hline
\multicolumn{2}{|c||}{\multirow{2}{3.8cm}{\scriptsize{\textbf{Strongest Single Beam Power RMS Delay Spread Over All Locations with Signal\footnotemark[2]}}}} & \multicolumn{2}{c|}{\textbf{LOS}} & \multicolumn{2}{c|}{\textbf{NLOS}} \tabularnewline \cline{3-6}
\multicolumn{2}{|c||}{} 																					& \textbf{$\bm{\mu}$ (ns)} 	& \textbf{$\bm{\sigma}$ (ns)}	& \textbf{$\bm{\mu}$ (ns)}		& \textbf{$\bm{\sigma}$ (ns)}	\tabularnewline \hline \hline
\multicolumn{2}{|c||}{\textbf{28 GHz (10.9$^\circ$ HPBW)}}	& 1.19\footnotemark[3] 							& 0.76\footnotemark[3] 						& 25.7 							& 40.2 							\tabularnewline \cline{2-6} 
\multicolumn{2}{|c||}{\textbf{73 GHz (7$^\circ$ HPBW)}}		& 1.16 							& 0.42 						& 7.1 							& 8.3 							\tabularnewline \hline
\end{tabular}
\end{center}
\end{table}
\footnotetext[2]{\textit{Strongest single beam power} refers to the unique antenna pointing angles in the azimuth and elevation plane at both the TX and RX for each location combination that resulted in the strongest received power.}
\footnotetext[3]{Mean and standard deviation values are calculated for five of six LOS locations at 28 GHz. The sixth LOS location had a larger than normal RMS delay spread (153.5 ns) as the TX and RX antennas were not properly aligned on boresight for that T-R separation distance, due to not searching for the best pointing angles.}

\section{Distance Extension Using Beam Combining}~\label{sec:CovExt}
Beamforming and beam combining techniques will be feasible for mmWave wireless and will improve link margin at cell edges, or equivalently offer, increased cell coverage beyond the estimated 200 m cell radius~\cite{Shu:CommMagDec14}. We quantify the distance extension by introducing a new concept of a \textit{distance extension exponent} (DEE), used to determine distance extension curves as a function of T-R separation distance by combining beams at the mobile handset.

Enabling distance extension through beam combining can be understood by considering two mean path losses $\PL_{\mathrm{(1\;beam)}}(d_1)$ at distance $d_1$ for the single best beam and $\PL_{\mathrm{(multibeam)}}(d_2)$ at distance $d_2$ for combining beams. Both path loss values are equivalent such that:
\begin{equation}\label{eq4}
\PL_{\mathrm{(1\;beam)}}(d_1) = \PL_{\mathrm{(multibeam)}}(d_2)
\end{equation} 
Below, the DEE derivation is based on a close-in free space reference distance, $d_0 = 1$ m, as this provides a standard for all measurements and models standardized to a 1 m reference. The distance extension exponent can be determined by solving for $d_2$ in terms of $d_1$, Eq.~(\ref{eq4}) remains valid for all $(d_1,d_2)$ pairs. Given the two path loss exponents (PLE) $n_1$ (single best beam) and $n_2$ (multibeam) (with respect to a free space reference distance $d_0 = 1$ m), we can estimate $\PL_{\mathrm{(1\;beam)}}(d_1)$ and $\PL_{\mathrm{(multibeam)}}(d_2)$ using the close-in free space reference distance path loss models in the following way:
\begin{equation}\label{eq5}
\PL_{\mathrm{(1\;beam)}}(d_1) =  \PL_{\mathrm{FS}}(d_0) + 10 n_1 \log_{10}\left(      \frac{d_1}{d_0}\right)
\end{equation}
\begin{equation}\label{eq6}
\PL_{\mathrm{(multibeam)}}(d_2) =  \PL_{\mathrm{FS}}(d_0) + 10 n_2 \log_{10} \left(\frac{d_2}{d_0} \right) 
\end{equation}
where $\PL_{\mathrm{FS}}(d_0)$ is the free space path loss at the standardized reference distance $d_0 = 1$ m. After substituting~(\ref{eq5}) and~(\ref{eq6}) into~(\ref{eq4}), we obtain the relationship between $d_2$ and $d_1$ such that~(\ref{eq4}) remains valid:
\begin{equation}
\left(d_2 \right)=    \left(d_1\right)  ^{\frac{n_1}{n_2}},\;(n_1\geq n_2,\:\mathrm{\textbf{always}})
\end{equation}
where $\frac{n_1}{n_2}$ defines the DEE.  Table~\ref{tbl:CovExt} provides empirical values for the DEE under different beam combining scenarios, where $n_1$ describes the PLE under the single best beam, and $n_2$ describes the PLE for coherently or non-coherently combining multiple beams~\cite{Shu:CommMagDec14}. Coherent combining is done by summing the square root of strongest received powers, in Volts and squaring the sum, resulting in Watts, whereas non-coherent combining simply adds the strongest received powers together in Watts, and both methods are performed during post-detection~\cite{Shu:CommMagDec14,SR:IMS14}. For example, a maximum cell radius of 200 m for the single best beam can be extended to 448 m when combining the four best beams coherently (DEE = 1.152), extending the coverage distance by a factor of 2.24 as seen in Eq.~(\ref{eq:DEE1}).  
 
\begin{figure}
	\begin{center}
		\includegraphics[width=0.5\textwidth]{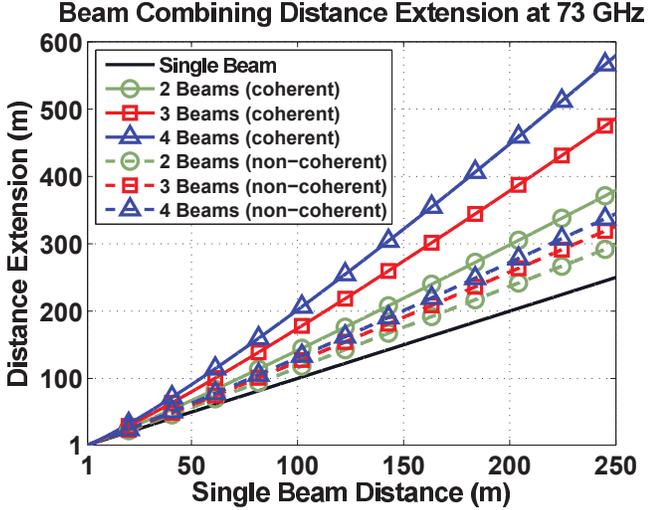}
		\caption{Beam combining distance extension curves at 73 GHz for coherently and non-coherently combining the best beams compared the single best beam. 28 GHz beam combining distance extension follows the same trend as the 73 GHz data.}\label{fig:CEF}
	\end{center}
\end{figure}

The effective \textit{distance extension factor} (DEF) for a given T-R separation distance $d_1$ (PLE  = $n_1$) when combining the best four beams compared to the single best beam, that results in $d_2$ (PLE = $n_2$) is computed as follows:
\begin{equation}
\mathrm{DEF}(d_1,d_2) = \frac{d_2-d_1}{d_1}+1
\end{equation}
or equivalently using the DEE:
\begin{equation}
\mathrm{DEF}(d_1,\mathrm{DEE}) = \left[d_1\:^{\left(\mathrm{DEE-1}\right)}\right]
\end{equation}
\begin{equation}\label{eq:DEE1}
\mathrm{DEF}(200,1.152) = \left[200^{\left(1.152-1\right)}\right] = 2.24
\end{equation}
The DEE can be used to determine the increased distance for a user experiencing the same path loss for beam combining compared to the single best beam. Fig.~\ref{fig:CEF} displays the distance extension for coherently and non-coherently combining the best beams compared to the single best beam at 73 GHz. The 28 GHz distance extension curves follow the same trend as shown in Table~\ref{tbl:CovExt}. The use of a 1 m free space reference distance allows a standard propagation model to be used and allows such comparisons in system design.
\begin{table}
\renewcommand{\arraystretch}{1.1}
\begin{center}
\caption{NLOS beam combining distance extension exponents (DEE) when considering the combination of the strongest beams compared to the single strongest beam and the corresponding PLEs with respect to a 1 m free space reference distance $d_0$. A comparison in distance extension is also given for a path loss observed at a distance of 200 m for the single best beam and the distance extension when combining the best beams for which the same path loss would be observed.}~\label{tbl:CovExt}
\begin{tabular}{| >{\centering\arraybackslash}m{0.85cm} | >{\centering\arraybackslash}m{0.7cm} |>{\centering\arraybackslash}m{0.6cm} |>{\centering\arraybackslash}m{0.6cm} | >{\centering\arraybackslash}m{0.75cm} | >{\centering\arraybackslash}m{0.6cm} | >{\centering\arraybackslash}m{0.6cm} | >{\centering\arraybackslash}m{0.75cm} |}\hline
\multicolumn{8}{|c|}{\textbf{Beam Combining PLE and Distance Extension Exponent ($\bm{d_0}$ = 1 m)}} \\ \hline
\textbf{Freq.} & & \multicolumn{6}{c|}{\textbf{PLE (Over all angles) = 4.556}} \\ \hline
\multirow{6}{*}{\textbf{28 GHz}} & \textbf{10.9$^\circ$ HPBW} & \multicolumn{3}{c|}{\textbf{Coherent}} & \multicolumn{3}{c|}{\textbf{Non-Coherent}} \\ \cline{2-8}
& \textbf{Beams} 	& \textbf{PLE} 	& \textbf{DEE} 				& \textbf{$\bm{d_2}$ if}		& \textbf{PLE} 		& \textbf{DEE} 			& \textbf{$\bm{d_2}$ if} 		\\ \cline{2-4} \cline{6-7}
& 1 				& 3.812 & - 	& \textbf{$\bm{d_1}$ = 200 m}& 3.812 				& - 				& \textbf{$\bm{d_1}$ = 200 m}  							\\ \cline{2-8}
& 2 				& 3.548 & 1.074 & 296 m						& 3.692 				& 1.033				& 238 m												\\ \cline{2-8}
& 3 				& 3.406 & 1.119 & 376 m						& 3.631 				& 1.050 			& 261 m												\\ \cline{2-8}
& 4 				& 3.307 & 1.153 & 450 m						& 3.591 				& 1.062				& 278 m												\\ \hline
& & \multicolumn{6}{c|}{\textbf{PLE (Over all angles) = 4.687}} \\ \cline{2-8}
\multirow{6}{*}{\textbf{73 GHz}} & \textbf{7$^\circ$ HPBW} & \multicolumn{3}{c|}{\textbf{Coherent}} & \multicolumn{3}{c|}{\textbf{Non-Coherent}} \\ \cline{2-8}
& \textbf{Beams} 	& \textbf{PLE} 	& \textbf{DEE} 				& \textbf{$\bm{d_2}$ if}		& \textbf{PLE} 		& \textbf{DEE} 			& \textbf{$\bm{d_2}$ if} 		\\ \cline{2-4} \cline{6-7}
& 1 				& 3.728 & - 	& \textbf{$\bm{d_1}$ = 200 m} 	& 3.728 				& - 				& \textbf{$\bm{d_1}$ = 200 m}  							\\ \cline{2-8}
& 2 				& 3.466 & 1.076 & 300 m						& 3.613 				& 1.032 			& 237 m												\\ \cline{2-8}
& 3 				& 3.327 & 1.121 & 380 m						& 3.557 				& 1.048				& 258 m												\\ \cline{2-8}
& 4 				& 3.235 & 1.152 & 448 m						& 3.523 				& 1.058				& 272 m												\\ \hline
\end{tabular}
\end{center}
\end{table}

\section{3-D Omnidirectional RMS Delay Spreads}~\label{sec:3Drms}
\subsection{Recovering Omnidirectional RMS Delay Spreads}
\begin{figure}
	\begin{center}
		\includegraphics[width=3.5in]{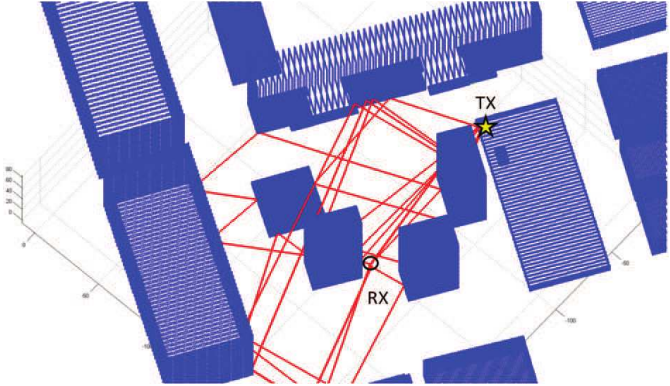}
		\caption{A 3-Dimensional view of the Downtown Manhattan area obtained from the MATLAB-based 3-D ray-tracer. The rays which leave the TX and successfully arrive at the  {\color{black}measured} RX are shown in red, and represent multipath signal trajectories in a dense UMi. The TX was located on the rooftop of the Coles Sports Center 7 m above ground (yellow star), and the RX was located 113 m away, 1.5 m above ground (black circle)~\cite{MR:GCOM14}.}\label{fig:RayTracedMap}
	\end{center}
\end{figure}

\begin{figure}
	\begin{center}
		\includegraphics[width=3.5in]{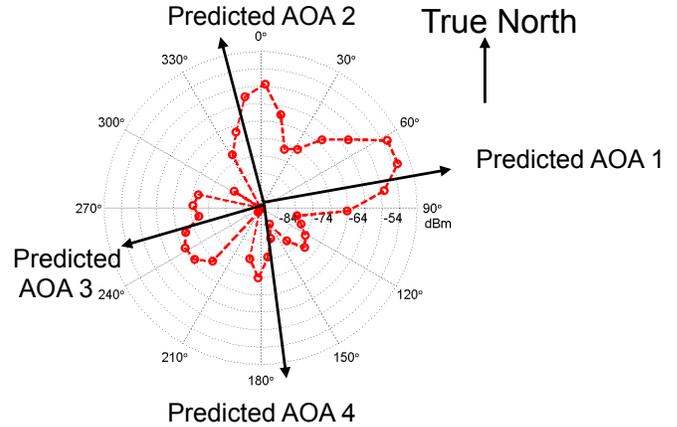}
			\caption{Measured azimuthal distribution of total received power (dBm units), also referred to as a polar plot, showing the predicted AOAs (black arrows) using 3-D ray-tracing at {\color{black}the} measured Manhattan RX location  {\color{black}shown in Fig.~\ref{fig:RayTracedMap}}. The center of the plot corresponds to the RX location. The RX and TX antennas both had 24.5 dBi of gain with $10.9^{\circ}$ (in azimuth) and $8.6^{\circ}$ (in elevation) 3 dB beamwidths, and the RX azimuth $0^{\circ}$ mark points to the True North bearing direction~\cite{MR:GCOM14}.}\label{fig:Polar}
	\end{center}
\end{figure}

Omnidirectional PDPs were recovered at a majority of the measured RX locations from the directional measurements using a MATLAB-based 3-D ray-tracing package developed at NYU to emulate electromagnetic propagation, and to predict absolute time delays of propagating MPCs. Table~\ref{tbl:synthPDP} shows the number of locations where omnidirectional PDPs were synthesized compared to the total number of measured locations, as not all locations were able to be synthesized since the database did not agree with our observations. An $800 \times 800$  $\mathrm{m}^2$ area was modeled in Google SketchUp, allowing 3-D environment-specific modeling with an accuracy of 5 m. Fig.~\ref{fig:RayTracedMap} shows a typical ray-traced synthesized RX location where viable propagation paths are shown in red. The corresponding measured power azimuth spectrum is shown in Fig.~\ref{fig:Polar} along with the ray-tracing predictions for the strongest angles of arrival~\cite{MR:GCOM14}. Our ray-tracer predicted up to four strongest measured angles at each RX location with an accuracy of $\pm 20^{\circ}$, which was enough to estimate the absolute propagating time delays of the first arriving multipath components at each of the strongest measured angles. We note the ray-tracer was unable to predict angles that received weakly scattered energy as a result of the coarseness of the database used.

We synthesized absolute timing omnidirectional PDPs using the ray-tracing predictions with the measured data. The PDPs corresponding to the strongest measured azimuth and elevation angles were matched up with the corresponding shortest predicted absolute propagation distances resulting from the ray-tracing simulations.  The excess delay PDPs from the strongest measured angles were then superimposed upon an absolute propagation time delay axis, where each PDP was appropriately shifted and added in time using the ray-tracing predicted absolute propagation time delay (obtained from dividing the predicted distance by the speed of light in free space)~\cite{MR:GCOM14}. {\color{black}The strongest multipath components propagating from TX to RX were assumed to follow specular reflection paths and result in prominent angles, which could be simulated in the ray-tracing environment using the ``Law of Reflection", a reasonable assumption in order to pair up strong measured angles and ray-tracing predictions.} The resulting synthesized 3-D omnidirectional PDPs for measured RX locations were subsequently analyzed to extract the RMS delay spreads~\cite{ICC2015:SR}. 
\begin{table}
\renewcommand{\arraystretch}{1}
\begin{center}
\caption{Number of locations where omnidirectional PDPs were synthesized via ray-tracing.}~\label{tbl:synthPDP}
\begin{tabular}{|M{4cm}|M{1.5cm}|M{1.5cm}|}
\hline
 & \textbf{28 GHz} & \textbf{73 GHz} \tabularnewline \hline
\textbf{LOS locations synthesized for omnidirectional PDPs} & \textbf{3} out of 6 & \textbf{5} out of 5 \tabularnewline \hline
\textbf{NLOS locations synthesized for omnidirectional PDPs} & \textbf{13} out of 20 & \textbf{19} out of 25 \tabularnewline \hline
\end{tabular}
\end{center}
\end{table}

\subsection{Millimeter-Wave Omnidirectional RMS Delay Spreads}
\begin{figure}
    \centering
 \includegraphics[width=3.5in]{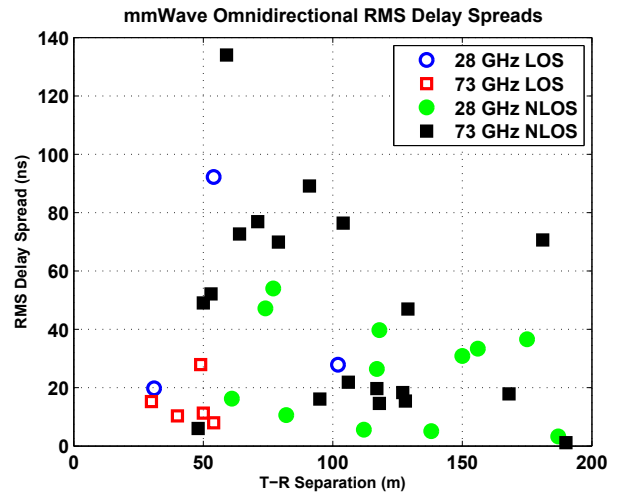}
    \caption{Omnidirectional RMS delay spreads as a function of T-R separation distance obtained from synthesized absolute timing PDPs measured at 28 GHz and 73 GHz for LOS and NLOS channels in Downtown Manhattan, a dense UMi environment.}
    \label{fig:RMS}
\end{figure}

Fig.~\ref{fig:RMS} shows the 28 GHz and 73 GHz omnidirectional RMS delay spreads as a function of T-R separation distance in both LOS and NLOS environments, obtained from absolute timing omnidirectional PDPs. The mean LOS RMS delay spreads were 46.6 ns and 14.6 ns at 28 GHz and 73 GHz, respectively, and the corresponding NLOS values were 40.9 ns and 45.7 ns. These values depend significantly on the ray-tracing simulation results, required to recover the omnidirectional profiles. We also note that there are only three locations with synthesized omnidirectional RMS delay spreads for the 28 GHz LOS case. We again note that the 28 GHz measurements did not search for the angles that resulted in the strongest received power and did not measure all locations with antennas aligned on boresight.  The RMS delay spread appears to  {\color{black}slightly} decrease as the T-R separation distance increases, where MPCs experience more reflections to arrive at the receiver. Finally, the 73 GHz mean omnidirectional NLOS RMS delay spreads are slighly larger than at 28 GHz, indicating that the received energy is more spread out in time, implying a more pronounced diffuse scattering effect at 73 GHz.
\begin{table}
\renewcommand{\arraystretch}{1.1}
\centering
\caption{Mean and standard deviation of the 28 GHz and 73 GHz omnidirectional RMS delay spreads obtained from absolute timing PDPs in LOS and NLOS environments in Downtown Manhattan.}
\begin{tabular}{|c|c|c|c|c|c|}
\hline
\multicolumn{2}{|c|}{\multirow{2}{*}{\textbf{Omni. RMS Delay Spread}}} & \multicolumn{2}{c|}{\textbf{LOS}} & \multicolumn{2}{c|}{\textbf{NLOS}} \tabularnewline \cline{3-6}

\multicolumn{2}{|c|}{}	& \textbf{ $\bm{\mu}$ (ns)} & \textbf{$\bm{\sigma}$ (ns)} & \textbf{ $\bm{\mu}$ (ns)} & \textbf{$\bm{\sigma}$ (ns)} \tabularnewline \hline

\multirow{2}{*}{Frequency} & 28 GHz & 46.6  & 39.7 & 40.9 & 57.0 \tabularnewline \cline{2-6}

				   & 73 GHz & 14.6 & 7.9 & 45.7 & 35.5 \tabularnewline \hline	
\end{tabular}
\label{tbl:41}
\end{table}

\begin{figure}
    \centering
 \includegraphics[width=3.5in]{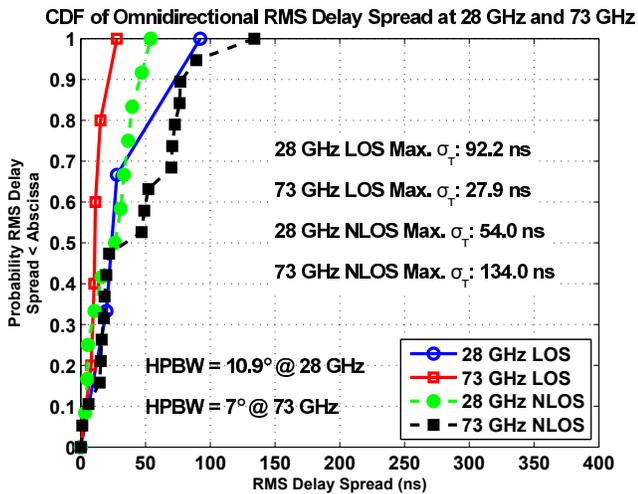}
    \caption{CDFs of the omnidirectional RMS delay spreads obtained from synthesized absolute timing PDPs measured at 28 GHz and 73 GHz in LOS and NLOS environments in Downtown Manhattan, a dense UMi environment.}
    \label{fig:RMSCDF}
\end{figure}

Fig.~\ref{fig:RMSCDF} shows the LOS and NLOS CDFs for the 28 GHz and 73 GHz omnidirectional RMS delay spreads. We observe that 50\% of all RMS delay spreads lie below 25 ns at both 28 GHz and 73 GHz, and that 90\% of all RMS delay spreads lie below 80 ns and 89 ns at 28 GHz and 73 GHz, respectively, illustrating more diffuse scattering at 73 GHz, as the received energy is more spread out in time delay.
\section{Conclusion}~\label{sec:conc}
This paper presented directional and omnidirectional RMS delay spread statistics obtained from outdoor mmWave UWB propagation measurements using highly directional steerable horn antennas, and presented a new system design parameter. Distance extension with beam combining using a \textit{distance extension exponent}, is useful in performing system-wide simulations. The DEE is made possible through the use of a 1 m free space reference distance for all LOS and NLOS path loss models, as described in~\cite{TSR:mmWave}. Directional measurements over all unique pointing angles resulted in mean RMS delay spreads of 28.8 ns and 13.9 ns in LOS at 28 GHz and 73 GHz, respectively, as seen in Fig.~\ref{fig:Drms} and Table~\ref{tbl:RMSdelay}. The directional mean RMS delay spreads were 17.4 ns and 11.1 ns in NLOS environments at 28 GHz and 73 GHz, respectively, showing that 28 GHz has larger RMS delay spread than 73 GHz. Measured NLOS directional mean RMS delay spreads at angles with the strongest received power are 25.7 ns and 7.1 ns at 28 GHz and 73 GHz, respectively, a great reduction over NLOS omnidirectional RMS delay spreads as seen in Table~\ref{tbl:41} where means of 40.9 ns and 45.7 ns were measured, respectively. The mean RMS delay spreads found at 28 GHz and 73 GHz in NLOS environments compare well with typical RMS delay spreads found at 59 GHz that are at or lower than 20 ns~\cite{Lovnes94:1, Smulders}. Future mmWave coverage distance can also be extended significantly through beam combining, by a factor of more than 2.2 from 200 m when combining the four strongest beams over the single best beam at 73 GHz in a NLOS environment using a DEE. The synthesized omnidirectional RMS delay spreads are much larger than at arbitrary pointing angles for both LOS and NLOS, indicating that directional mmWave systems will be more useful for avoiding ISI and thus improving signal quality and throughput. These results illustrate that mmWave UWB communications systems can be exploited with directional antennas at the TX and RX to achieve greater system performance.

\bibliographystyle{IEEEtran}
\bibliography{ICC_2015_bib}
\end{document}